\documentclass[aps,superscriptaddress,pra,twocolumn]{revtex4}
\usepackage{graphicx,amsmath,bbm,mathrsfs,amssymb,pst-all,bm,color}
\usepackage{epsfig}
\newcommand{\ket}[1]{\vert #1 \rangle}
\newcommand{\bra}[1]{\langle #1 \vert}
\newcommand{\dket}[1]{\vert #1 \rangle\rangle}
\newcommand{\dbra}[1]{\langle\langle #1 \vert}
 
\newcommand{\bmG}{\boldsymbol G}
\newcommand{\bmX}{\boldsymbol X}
\newcommand{\bmA}{\boldsymbol A} \newcommand{\bmM}{\boldsymbol M}
 \newcommand{\bmC}{\boldsymbol C}
\newcommand{\bmS}{\boldsymbol S} \newcommand{\bmR}{\boldsymbol R}
\newcommand{\bmu}{\boldsymbol u} \newcommand{\bmv}{\boldsymbol v}
\newcommand{\bmw}{\boldsymbol w}
\newcommand{\bmOmega}{{\boldsymbol \Omega}}
\begin{document}
\title{An effective method to estimate multidimensional Gaussian states}
\author{J.~\v{R}eh\'{a}\v{c}ek} 
\affiliation{Department of Optics, Palacky University, 
17. listopadu 50, 77200 Olomouc, Czech Republic}
\author{S.~Olivares}
\affiliation{CNISM UdR Milano Universit\`a, I-20133, Italia}
\affiliation{Dipartimento di Fisica dell'Universit\`a di Milano, I-20133, Italia}
\author{D.~Mogilevtsev}
\affiliation{Institute of Physics, Belarus National Academy of
Sciences, F.Skarina Ave. 68, Minsk 220072 Belarus}
\author{Z.~Hradil}
\affiliation{Department of Optics, Palacky University, 17. listopadu 50, 
77200 Olomouc, Czech Republic}
\author{M.~G.~A.~Paris}
\affiliation{Dipartimento di Fisica dell'Universit\`a di Milano, I-20133, Italia}
\affiliation{CNISM UdR Milano Universit\`a, I-20133, Italia}
\affiliation{Institute for Scientific Interchange Foundation, I-10133 Torino,
Italia}
\author{S. Fornaro}
\affiliation{Dipartimento di Scienze Fisiche Universit\`a "Federico
II", Complesso Universitario Monte Sant'Angelo, 80126 Napoli, Italy}
\affiliation{CNISM UdR Napoli, Complesso Universitario Monte
Sant'Angelo, 80126 Napoli, Italy}
\author{V. D'Auria}
\affiliation{Laboratoire Kastler Brossel, Universit\'e Pierre et
Marie Curie, Ecole Normale Sup\'erieure, CNRS, 4 place Jussieu, 75252 Paris, France}
\affiliation{Dipartimento di Scienze Fisiche Universit\`a "Federico
II", Complesso Universitario Monte Sant'Angelo, 80126 Napoli, Italy}
\author{A. Porzio}
\affiliation{CNISM UdR Napoli, Complesso Universitario Monte
Sant'Angelo, 80126 Napoli, Italy}
\affiliation{Dipartimento di Scienze Fisiche Universit\`a "Federico
II", Complesso Universitario Monte Sant'Angelo, 80126 Napoli, Italy}
\author{S. Solimeno}
\affiliation{Dipartimento di Scienze Fisiche Universit\`a "Federico
II", Complesso Universitario Monte Sant'Angelo, 80126 Napoli, Italy}
\affiliation{CNISM UdR Napoli, Complesso Universitario Monte
Sant'Angelo, 80126 Napoli, Italy}
\begin{abstract}
A simple and efficient method for characterization of multidimensional
Gaussian states is suggested and experimentally demonstrated. Our scheme
shows analogies with tomography of finite dimensional quantum states,
with the covariance matrix playing the role of the density matrix and
homodyne detection providing Stern-Gerlach-like projections. The major
difference stems from a different character of relevant noises: while
the statistics of Stern-Gerlach-like measurements is governed by binomial
statistics, the detection of quadrature variances correspond to $\chi^2$
statistics. For Gaussian and near Gaussian states the suggested method
provides, compared to standard tomography techniques, more stable and
reliable reconstructions.  In addition, by putting together
reconstruction methods for Gaussian and arbitrary states, we obtain a
tool to detect the non-Gaussian character of optical signals.
\end{abstract}
\date{}
\maketitle
\section{Introduction}
Gaussian states are building blocks of quantum information
processing with continuous variables. In fact, Gaussian states can be
generated and processed by means of linear operations. On the
other hand, successful implementation of quantum information
protocols requires efficient tools for the analysis and
characterization of the quantum states involved in the experiment
\cite{LNP}.  The standard approach to characterize optical
continuous variables states is to perform homodyne measurements
\cite{firsthomodyne} and reconstruct the measured quantum state
on a subspace of the infinitely dimensional space describing a
multi-mode quantum harmonic oscillator \cite{vogel}.  This
approach is rather general and, in principle, can be used for
reconstructing both Gaussian
\cite{firsthomodyne,squeezedcoherent,gaussian,CovMat,CMs} and non-Gaussian
\cite{onephoton,grg,polzik} states. On the other
hand, such a general procedure becomes inherently inefficient once the set
of possible states is restricted.  For example, the knowledge
about the Gaussian character of the measured state contributes a
lot of prior information about the measured  subject. This
information can be used for reducing the number of relevant
unknown parameters, making the reconstruction (data inversion)
much simpler, also avoiding problematic issues of the standard
approach, such as the rapid growth of reconstruction errors with
the size of the reconstruction space \cite{chyby,qt}.
\par
The main idea of the present contribution is to point out and 
exploit formal
analogies between the description of Gaussian states and that of 
finite-dimensional quantum states. Based on this analogy we put forward 
a simple and efficient method for reconstructing Gaussian
states from homodyne data. Since by definition the result of
Gaussian tomography is a Gaussian state, the quality of Gaussian
fit can also be used to assess how far is the measured state from
the family of Gaussian states, thus providing an operational definition
of Gaussianity for quantum states \cite{nonG}. This will be 
demonstrated upon the application of the method to homodyne data 
taken on states produced by an optical parametric oscillator close to
threshold \cite{solimeno}.
\par
The paper is structured as follows.  In Section \ref{REVIEW} we
review few basics facts related to Gaussian states and homodyne
detection. This will be used  in Section \ref{RECONSTRUCTION} for
the formulation of a reconstruction method based on the 
detection of rotated field-quadrature operator.  The theory is then 
generalized to multidimensional Gaussian states in
Section \ref{MULTIMODE}. Experimental results and data analysis 
are reported in in Sec.~\ref{EXPERIMENT}, whereas Section
\ref{outro} closes the paper with some concluding remarks.
\section{Homodyning Gaussian states}
\label{REVIEW}
For the sake of simplicity we start by formulating the problem of
reconstruction for single mode Gaussian field $\varrho$, with 
Wigner function given by
\begin{equation}\label{Wigner}
  W(\bmX) =
\frac{\exp\left\{ - \frac12
{ (\bmX-\overline{\bmX})^T  \bmG^{-1} (\bmX-\overline{\bmX}) }
\right\}
}{
2\pi \sqrt{{\rm Det}[\bmG]}}\: ,
\end{equation}
with $\bmX^T = (x,y)$ and covariance matrix $\bmG$ given by
\begin{equation}\label{G-matrix}
\bmG \equiv
\frac12
\begin{pmatrix}
  2 (\Delta X)^2 & \{ \Delta X, \Delta Y   \} \\
  \{ \Delta X, \Delta Y   \}& 2(\Delta P)^2
\end{pmatrix}
\ge  \bmOmega \equiv
\frac{1}{2}
\begin{pmatrix}
 0 & -i \\
i  & 0
\end{pmatrix}\,.
\end{equation}
where $X = (a + a^\dag)/\sqrt{2}$ and $Y = i(a^\dag-a)/\sqrt{2}$,
are the quadrature operators, $a$ and $a^{\dag}$ being mode operators.
If the homodyne detector is set to measure the quadrature
$X(\theta) = X \cos\theta + Y \sin\theta$,
the positive operator valued measure (POVM) $\Pi_\eta(x,\theta)$
associated with its realistic measurement is the Gaussian convolution 
of the quadrature projectors,
\begin{equation}
\Pi_\eta(x,\theta) = \int \frac{dy}{\sqrt{2\pi\delta_\eta^2}} \:
\exp\left\{-\frac{(y-
\frac{x}{\sqrt{\eta}})^2}{2\delta_\eta^2}\right\} \:
|y\rangle_\phi{}_\phi\langle y|,
\end{equation}
where $\delta_\eta^2 = (1-\eta)/2\eta$, and $\eta$ is the quantum
efficiency of the involved (linear) photodetectors.
In the Fock basis we have
\begin{equation}
|y\rangle_\phi = \left(\frac1\pi \right)^{1/4} e^{-\frac12 y^2}
\sum_{k=0}^\infty \frac{H_k(y)}{2^{k/2} \sqrt{k!}}  e^{-ik\phi}
|k\rangle\,,
\end{equation}
and the distribution of homodyne outcomes is given by
\begin{align}
p_\eta(x,\theta) &= {\rm Tr}\left[\varrho\: \Pi_\eta(x,\theta)\right]\\
 &= \frac{\exp\left\{
-\left(x - \left[{\bmR}_{\theta} \overline{\bmX}\right]_{1}\right)^2/
(2\,\sigma^2_\eta) \right\}
}{
\sqrt{ 2\pi\, \sigma^2_\eta}}
\,, \label{p_theta}
\end{align}
where $\sigma^2_\eta= \eta(M_{22} {\rm Det}[\bmG] + \delta_\eta^2)$, 
$M_{22} = [\bmM]_{22}$, with $\bmM = {\bmR}_{\theta}\,
\bmG^{-1} \bmR_{- \theta}$, and the rotation matrix is defined as
$$ \bmR_{\theta} =
\begin{pmatrix}
  \cos \theta & \sin \theta \\
  -\sin \theta& \cos \theta
\end{pmatrix}\,.
$$
\section{Estimation of covariance matrix}
\label{RECONSTRUCTION}
In this Section, we address the problem how to
estimate efficiently the covariance  matrix $\bm G$. For the sake
of simplicity let us set the coherent part of the signal to zero
and $\eta = 1$. Introducing a unit vector 
$\bra{\bmu} = (\cos{\theta }, \sin{\theta})$ parametrized by
the phase of the local oscillator and noticing that
$\bra{\bmu} \bmG \ket{\bmu}=M_{22} {\rm Det}[\bmG]$,
the sampled probability density (\ref{p_theta})
may be conveniently expressed in the form
\begin{equation}\label{sampled1}
p(x,\theta) = \frac{1} {\sqrt{ 2\pi\,\bra{\bmu} \bmG \ket{\bmu}}}
\exp\left\{-\frac{x^2}{2\, \bra{\bmu} \bmG \ket{\bmu}} \right\}\,.
\end{equation}
There are several striking similarities between the reconstruction of
spin $1/2$ states and covariance matrices. Indeed, in both cases the
state is described by $2\times2$ matrices. The density matrix of
the spin $1/2$ state is a hermitian, unit trace, semi-positive 
matrix, thus leaving three free
parameters for its full description. The covariance matrix is real,
symmetric matrix, again fully described by three parameters,
constrained to the relation (\ref{G-matrix}). 
Projections of a spin density matrix are conveniently 
sampled in a Stern-Gerlach experiment.
Similarly, detections of quadrature variables
in a homodyne experiment provide sampling of 
quadrature variances representing projections of a covariance 
matrix.
Indeed, estimated matrix $\bmG$ appears only in the variance of
distribution (\ref{sampled1}). Single detected event do not say
too much, but repeated detections do. Provided that homodyne
detection is repeated  $n$ times for the same setting,  $ x_i$
being the detected results, the likelihood of $\bmG$ reads
\begin{equation}\label{likelihood}
    {\cal L}(\bmG|\{ x_i \}) \propto
    \frac{1}{(2\pi\,\bra{\bmu} \bmG \ket{\bmu})^{n/2}}
\exp\left\{-\frac{ \sum_i x_i^2}{2\, \bra{\bmu} \bmG \ket{\bmu}}
\right\}.
\end{equation}
Consider now the statistics of the random variable  
$$ y = \sum_kx_i^2,$$ which is given by 
\begin{equation}\label{distribution}
P_{G}(y) = \int d x_1 \ldots dx_n \delta(y- \sum_i x_i^2) {\cal
L}(\bmG|\{ x_i \}).
\end{equation}
It is easy to see that for Gaussian states the fluctuations of 
$y$ variable are governed by the well-known $\chi^2$ distribution
\begin{equation}\label{chi}
 P_{G}(y) =   \frac{2^{-n/2}}{\Gamma(n/2)}
\frac{y^{n/2- 1}}{\big[ \sigma^2(\bmG) \big]^{n/2}} 
\exp\left\{ - \frac{y}{2\, \sigma^2(\bmG) }
\right\}.
\end{equation}
Upon maximizing the likelihood \eqref{chi}, the quadrature variance, 
which plays the role of a projection of the covariance 
matrix, $\sigma^2(\bmG) = \bra{\bmu} \bmG \ket{\bmu}$, 
is estimated as
\begin{equation}\label{variance1}
    \sigma^2(\bmG) = \frac{1}{n}\sum_i x_i^2.
\end{equation}
This establishes a formal analogy between estimations of spin states
and Gaussian covariance matrices. In the former case the probability
of finding the spin ``up'' is sampled
by the number of particles deflected upwards in a Stern-Gerlach apparatus.
Similarly, for Gaussian states the variance of
a quadrature distribution  (i.e. a projection of the 
covariance matrix) is sampled by properly normalized sum of 
squares of the detected quadrature values.
\par
Having mentioned similarities between the two problems let us also
identify three important differences between the estimation
of a spin $1/2$ system and estimation of a Gaussian state:
\begin{enumerate}
\item[(i)] the two problems have different underlying statics: 
binomial statistics for yes-no spin data and $\chi^2$ distribution 
for sampled variances;
\item[(ii)] there are slightly different constraints on the estimated 
quantities: semi-positivity of density matrix for spins and 
uncertainty relations constraint, stated in the form of the matrix 
inequality \eqref{G-matrix}, for Gaussian states;
\item[(iii)] the measured quantities have different nature: for spins
the sampled probabilities are \textit{expectation} values of Stern-Gerlach 
outcomes whereas for Gaussian states we need sampled variances of the
homodyne data, {\em i.e} the information is contained in the data
\textit{noise}.
\end{enumerate}
Notice that properties (ii) and (iii) will be important for the discussion 
of our main result below.
\par
In order to get a unique reconstruction of the measured Gaussian state,
quadrature measurement should be repeated with different settings of
phases $\theta_h$ of the strong local oscillator. Let us assume that
each quadrature $X(\theta_h)$ is sampled $n_h$ times with the results 
$x_{h,1},\,x_{h,2},\ldots x_{h, n_h}$, and denote $y_h=\sum_k x_{h,k}^2$.
The corresponding log-likelihood reads
\begin{equation}
\label{logLik}
  \log {\cal L}(\bmG)  = -\frac{1}{2} \sum_h n_h
\log \sigma_h^2(\bmG)  -
\sum_h \frac{y_h}{2\, \sigma_h^2(\bmG) }\,,
\end{equation}
where $ \sigma_h^2(G) = \bra{\bmu_h} \bmG \ket{\bmu_h}$.
A reconstruction of the signal covariance matrix 
is then obtained by maximizing the likelihood function
subject to the constraint $\bf G\ge\Omega$.
Let us first consider the extremal equation for the 
covariance matrix. It is easily found by taking a derivative of 
$\log {\cal L}(\bmG)$ with respect to matrix $\bmG$. This yields
\begin{equation}
\label{ext-eqG} R \bmG =  D \bmG,
\end{equation}
where
\begin{align}
D &=  \sum_h \frac{n_h}{\langle \bmu_h | \bmG | \bmu_h \rangle }|
\bmu_h \rangle \langle \bmu_h | \,, \\
R &=
\sum_h \frac{y_h}{\langle \bmu_h | \bmG | \bmu_h \rangle^2 }
| \bmu_h \rangle \langle \bmu_h |.\label{operR}
\end{align}
As there is again a direct analogy between the extremal equation for
$\bm G$ and the corresponding extremal equation for the maximum-likely
spin state, methods developed for generic quantum estimation can be
employed, with some caution, to solve the extremal  equation
(\ref{ext-eqG}) by iterations \cite{mylvovsky}.  A form suitable to
iterations can be found noting that by Hermicity of $\bmG$, $R$ and $D$
both $\bmG = D^{-1} R \bmG$ and $\bmG = \bmG R D^{-1}$ simultaneously
hold and may be combined into a single extremal matrix equation
\begin{equation}
\label{solution} \bmG = D^{-1} R \bmG R D^{-1},
\end{equation}
which is our main formal result. This matrix equation may be iterated
starting e.g. with the covariance matrix of the vacuum state.  In
addition the form of Eq.~\eqref{solution} guarantees that
semi-positivity, $\bmG\ge 0$, of the covariance matrix is preserved
after each iteration step.  \par 
Some discussion of this algorithm is now in
order.  As we have already mentioned above condition $\bmG\ge 0$ would
be typical for density matrices. Covariance matrices obey a more
complicated relation $\bm G\ge \bm \Omega$.  Though {\it in principle}
the algorithm \eqref{solution} may converge to a covariance matrix
violating Heisenberg uncertainty relations, we will present physical
arguments showing that {\it in practice} this almost never happens.
\par
First, let us discuss the estimation of a spin state from noisy data.
Ignoring for a while the constraint $\rho\ge 0$, the probability that
the best fit of measured data is provided by a non physical matrix
depends on the distance of $\rho$ from the boundary of the convex set of
density matrices and also on the character of the noisy data.  When the true
state lies close to the boundary and/or the measured relative
frequencies differ a lot from the theoretical probabilities the chances
are high that the data are best explained by a ``density matrix'' having
at least one negative eigenvalue. In addition, states close to the
boundary are easy to produce e.g. by a projection. Hence the condition
$\rho\ge 0$ is an essential part of the spin state tomography, which
must be incorporated in any meaningful state reconstruction protocol.
\par
The situation is quite different in Gaussian state tomography where the
boundary of physical covariance matrices is made by minimum
uncertainty states, which rarely appear in practice.  In fact, in order 
to generate a minimum uncertainty state one has to eliminate {\em all} 
the sources of noise from the preparation procedure. In addition, 
variances (noise) rather then probabilities are fitted from data in Gaussian 
tomography. Hence, any additional noise, being of statistical origin or due 
to some imperfection in the measurement, will have the effect of moving the 
reconstructed state further away from the boundary.  For these reasons the 
condition $\bm G\ge \bm \Omega$ is not likely to be violated by $\bm G$
reconstructed from real data.  This justifies the use of the simple
reconstruction algorithm of Eq.~\eqref{solution}.
Cast in another way, a possible violation of the condition $\bm G\ge \bm
\Omega$ can be seen as a challenge for experimenters. Indeed, to obtain
such a data set, one must be able to prepare and control quantum states
lying very close to the minimum uncertainty states, and to measure them
with sufficient precision. This requirement is obviously much more
demanding then, for example, the generation and observation of squeezing
in one quadrature without paying attention to the excess noise
introduced in the conjugated observable.  In case of such an advanced
experiment the condition $\bm G\ge \bm \Omega$ can taken into account 
by means of a simple two-step strategy.  Provided $\bm G\ge \bm \Omega$ the
reconstructed covariance matrix  is physically sound and the search is
over.  In the opposite case a search for the maximum-likely covariance
matrix should be repeated among the set of minimal uncertainty states.
\section{Reconstruction of multi-mode Gaussian states}
\label{MULTIMODE}
The reconstruction method illustrated in the previous Section 
may be straightforwardly generalized to the problem of estimating
the covariance matrix of multi-mode Gaussian states. 
The strategy is the following. The modes will be
mixed by means of controlled beam splitter and the homodyne
detection will be done on the selected output mode. Provided that
covariance matrix  has been sampled by sufficient number of
projections, it can be reconstructed. In order
to apply the ML procedure developed above it is enough to show
how variances of the detected modes can be derived from the
generic covariance matrix. Without loss of generality the method
will be illustrated on the example of two-mode case 
since the extension of the model to higher dimensional case is
straightforward.
\par
Let $\bmG_{\rm in}$ be the $4 \times 4$ covariance matrix of a
two-mode Gaussian state with entries
\begin{align}\label{definice}
G_{kl} \equiv [\bmG_{\rm in}]_{kl} &=  \frac12 \langle \{Q_k,Q_l\} \rangle -
\langle  Q_l \rangle
\langle  Q_k \rangle
\end{align}
where
$\boldsymbol{Q}=(X_1,Y_1,X_2,Y_2)$, $\{Q_k,Q_l\} = Q_k Q_l + Q_l Q_k$,
and $X_k = \frac{1}{\sqrt{2}}(a_k + a^\dag_k)$ and
$Y_k = \frac{1}{i\sqrt{2}}(a_k - a^\dag_k)$ are the quadrature operators
of mode $k$. Moreover, one should have
\begin{equation}
\bmG_{\rm in} \ge \bmOmega \oplus \bmOmega =
\left(
\begin{array}{c|c}
\bmOmega & {\boldsymbol 0} \\
\hline
{\boldsymbol 0} & \bmOmega
\end{array}
\right)\,,
\end{equation}
${\boldsymbol 0}$ being the $2\times2$ null matrix.
If the two modes are mixed at a beam splitter with
transmissivity $\tau = \cos^2 \vartheta$, then the emerging two-mode state
is still Gaussian and its covariance matrix is given by:
\begin{equation}
\bmG_{\rm in} \to \bmG_{\rm out} =
\bmS_{\rm BS}(\vartheta)\, \bmG_{\rm in}\, \bmS_{\rm BS}^{T}(\vartheta)
\end{equation}
where
\begin{equation}
\bmS_{\rm BS}(\vartheta) = \left(
\begin{array}{c|c}
\cos\vartheta\, {\mathbbm 1}_2 & \sin\vartheta\, {\mathbbm 1}_2 \\
\hline
-\sin\vartheta\, {\mathbbm 1}_2 & \cos\vartheta\, {\mathbbm 1}_2
\end{array}
\right)
\end{equation}
is the symplectic transformation associated with the beam splitter and
${\mathbbm 1}_2$ is the $2 \times 2$ identity matrix. In the case of a
balanced beam splitter ($\vartheta = \pi/4 $) the two output modes,
say $b_1$ and $b_2$, correspond $b_1 = (a_1 + a_2)/\sqrt{2}$ and
$b_2 = (a_1 - a_2)/\sqrt{2}$, respectively.
In this case we can write the evolved covariance matrix as follows:
\begin{equation}
\bmG_{\rm out} = \left(
\begin{array}{c|c}
\bmA^{(+)} & \bmC \\
\hline
\bmC^T & \bmA^{(-)}
\end{array}
\right)
\end{equation}
with
\begin{align}\label{output:CM}
\bmA^{(\pm)} = \frac12  \left[ \bmG_1 + \bmG_2 \pm (\bmG_3 + \bmG_4) \right],\\
\bmC = \frac12  \left[ \bmG_2 - \bmG_1 + (\bmG_3 - \bmG_4) \right],
\end{align}
where:
\begin{align}\label{input:CM}
\bmG_{1} = \left(
\begin{array}{cc}
G_{11} & G_{12} \\
G_{12} & G_{22}
\end{array}
\right) \,,
&\quad \bmG_{2} = \left(
\begin{array}{cc}
G_{33} & G_{34} \\
G_{34} & G_{44}
\end{array}
\right)\,,\\
\bmG_{3} = \left(
\begin{array}{cc}
G_{13} & G_{14} \\
G_{23} & G_{24}
\end{array}
\right) \,,
&\quad \bmG_{4} = \bmG_{3}^{T}\,,
\end{align}
and the elements $G_{kl}$ are defined by Eq.~\eqref{definice}.  Matrices
$\bmG_1$ and $\bmG_2$ correspond to the covariance matrices of the
reduced single-mode input states, i.e., when the other one is neglected.
We have chosen the notation in such a way that $\bmA^{(+)}$ and
$\bmA^{(-)}$ are the covariance matrices of the {\em single-mode} states
corresponding to modes $b_1$ and $b_2$, respectively.  Homodyne
detection on mode $b_1$ can be simply described by extension of vector
$|\bmu \rangle$ into higher dimension as $ \dbra{\bmu} = (\cos \theta,
\sin \theta, 0,0)$; similarly, in the case of mode $b_2$ one has to use
the vector $ \dbra{\bmv} = (0,0,\cos \theta, \sin \theta)$.  Hence the
detected variance can be easily represented by expression  $ \dbra{\bmw}
\bmS_{\rm BS}(\vartheta)\,\bmG_{\rm in} \bmS_{\rm BS}^T(\vartheta)\,
\dket{\bmw}$, with $\bmw = \bmu, \bmv$.  Consequently, the
reconstruction can be done by means of the same reconstruction algorithm
as above considering index $h$ as multi-index for phase of homodyne
detection $\theta_k$ and the beam splitter transmissivity
$\tau=\cos^2\vartheta$:
\begin{equation}\label{quorum}
    |\bmu_h \rangle \rightarrow \bmS_{\rm BS}^T(\vartheta)\, \dket{\bmu_k}\,.
\end{equation}
In particular, two configuration of beam splitter are sufficient for
successful reconstruction of 2-mode covariance matrix, namely the
detection on the free (uncoupled) modes with $\bmS_{\rm BS}(0)= \mathbbm 1$
[corresponding to covariance matrices (\ref{input:CM})] and
mixing with symmetric  beam splitter [corresponding to covariance
matrices (\ref{output:CM})]. Taking into account the
efficiency, the extremal equation for covariance matrix reads
\begin{equation}
\label{generic_sigma} 
D(\bmw) \bmG =  R(\bmw) \bmG,
\quad\quad \bmw=\bmu,\bmv
\end{equation}
where 
\begin{align}
 D(\bmw) = \sum_{k, \vartheta}&
\frac{1}{ \dbra{\bmw_k} \bmS_{\rm BS}(\vartheta)\, \bmG
\bmS_{\rm BS}^T(\vartheta) \dket{\bmw_k} + \delta^2_{\eta}}
\nonumber\\ 
&\times \bmS_{\rm BS}^T(\vartheta)
\dket{\bmw_k} \dbra{\bmw_k} \bmS_{\rm BS}(\vartheta)
\end{align}
and
\begin{align}
 R(\bmw) = \sum_{k, \vartheta}&
\frac{x_{h,\vartheta}^2}{ \left[\dbra{\bmw_k} \bmS_{\rm BS}(\vartheta)\,
\bmG \bmS_{\rm BS}^T(\vartheta) \dket{\bmw_k} + \delta^2_{\eta} \right]^2}
\nonumber\\
&\times\bmS_{\rm BS}^T(\vartheta)
\dket{\bmw_k} \dbra{\bmw_k} \bmS_{\rm BS}(\vartheta)\,.
\end{align}
\section{Experiments and data analysis}
\label{EXPERIMENT}
\begin{figure}[h!]
\centerline{
\includegraphics[width=0.75\columnwidth]{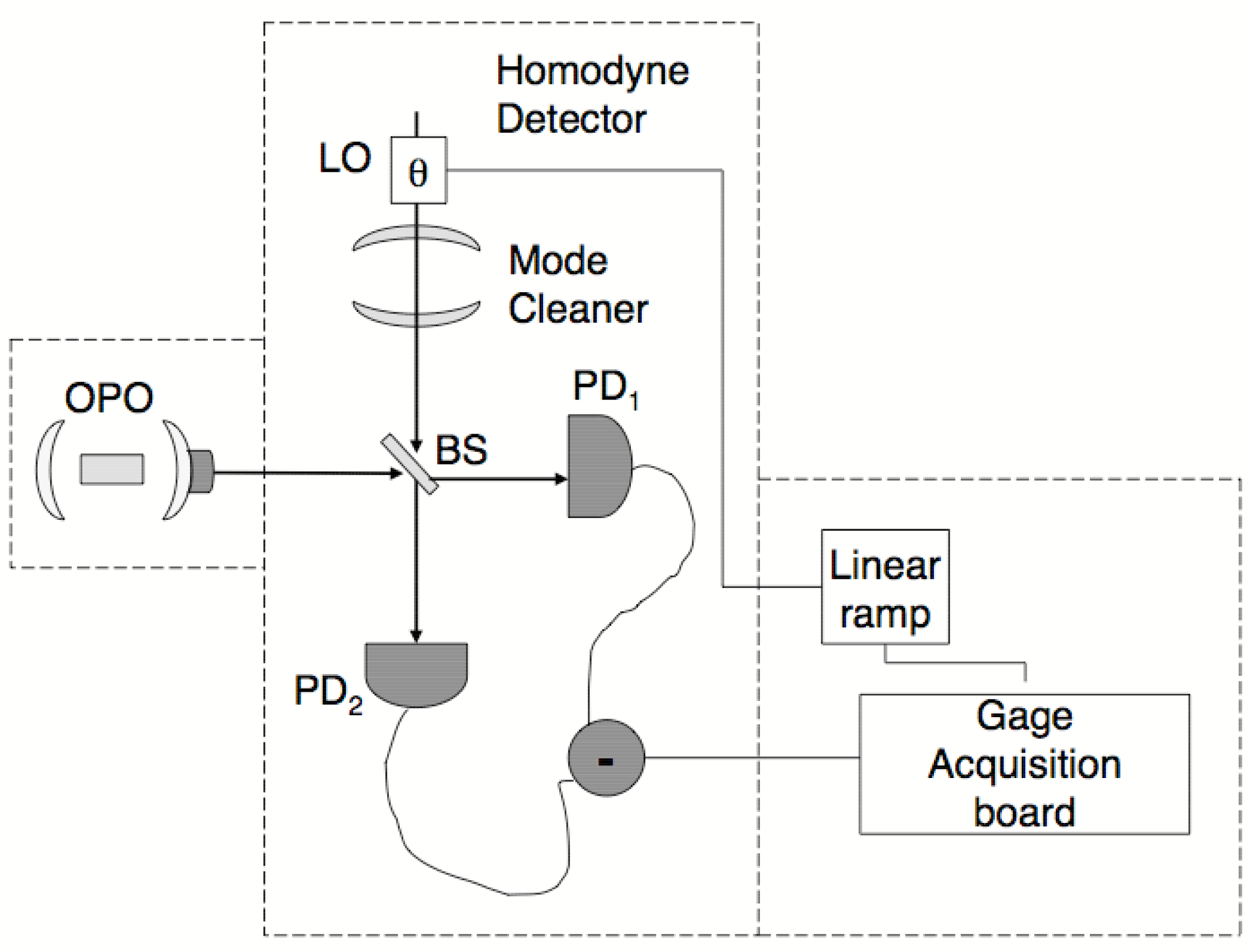}}
\caption{\label{f:exp} Experimental setup.}
\end{figure}

The method exposed in the previous section has been applied to homodyne
data samples acquired for two different kind of input state: a
(definitely Gaussian) reference vacuum state and slightly non Gaussian state, close
to squeezed thermal vacuum state, generated by an I-type optical
parametric oscillator (OPO) that operates close to the threshold.  In
the following, we give a brief description of the experimental setup \cite{solimeno}
with particular attention
payed to the homodyne detector and the data acquisition system.  The
setup can be divided into three distinct blocks: the state source, a
below threshold fully degenerated OPO, the detector, a quantum homodyne
and related electronics, and the, PC based, acquisition board (see
Fig.~\ref{f:exp}).
\par
The source is a below threshold fully degenerate OPO based on a type--I
LiNbO$_{3}$:MgO crystal and pumped at 532~nm \cite{APB2001}.  The
quantum homodyne detector, whose reliability has been proved in
different experiments \cite{solimeno,CovMat}, shows an
overall quantum efficiency of $\eta =0.88\pm 0.02$. To avoid any
influence on the statistics of the data, the electronic noise floor is
kept $\approx 15$~dB below the shot noise. Mode matching at the BS has
been accomplished by spatially filtering the local oscillator (LO) beam
by a mode cleaner cavity whose geometrical properties match the OPO's ones.
The relative phase between the LO and the beam exiting the OPO is scanned
by a piezo mounted mirror to which a linear voltage ramp is applied;
the ramp is adjusted so to have a 2$\pi$ phase span in the acquisition
time. The 2$\pi $ variation necessary for a full state reconstruction is
selected in the central region of the piezo range so to optimize the
linear response of the piezo stack.
\begin{figure}[h!]
\centerline{
\includegraphics[width=0.7\columnwidth]{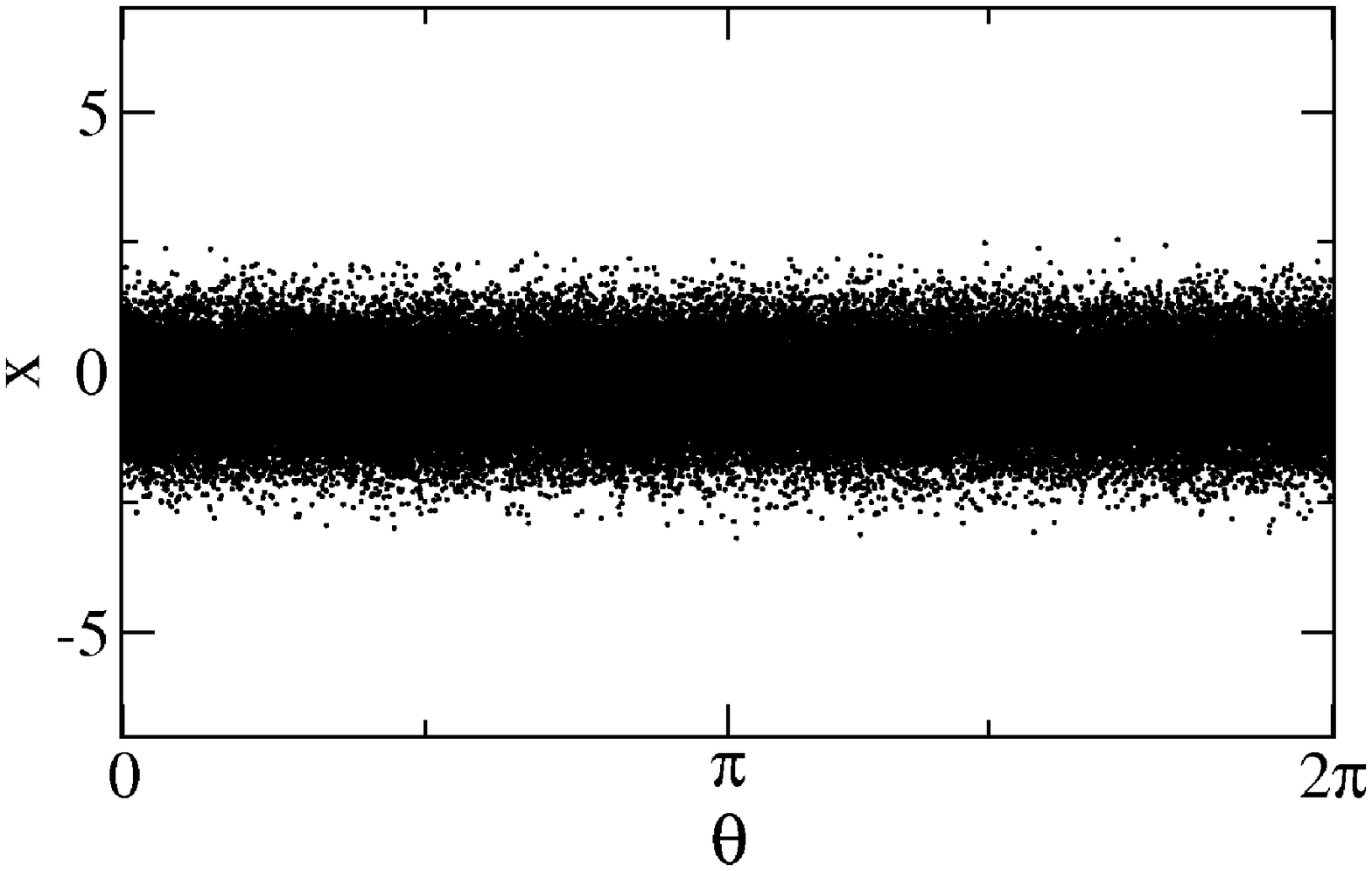}}
\centerline{\includegraphics[width=0.7\columnwidth]{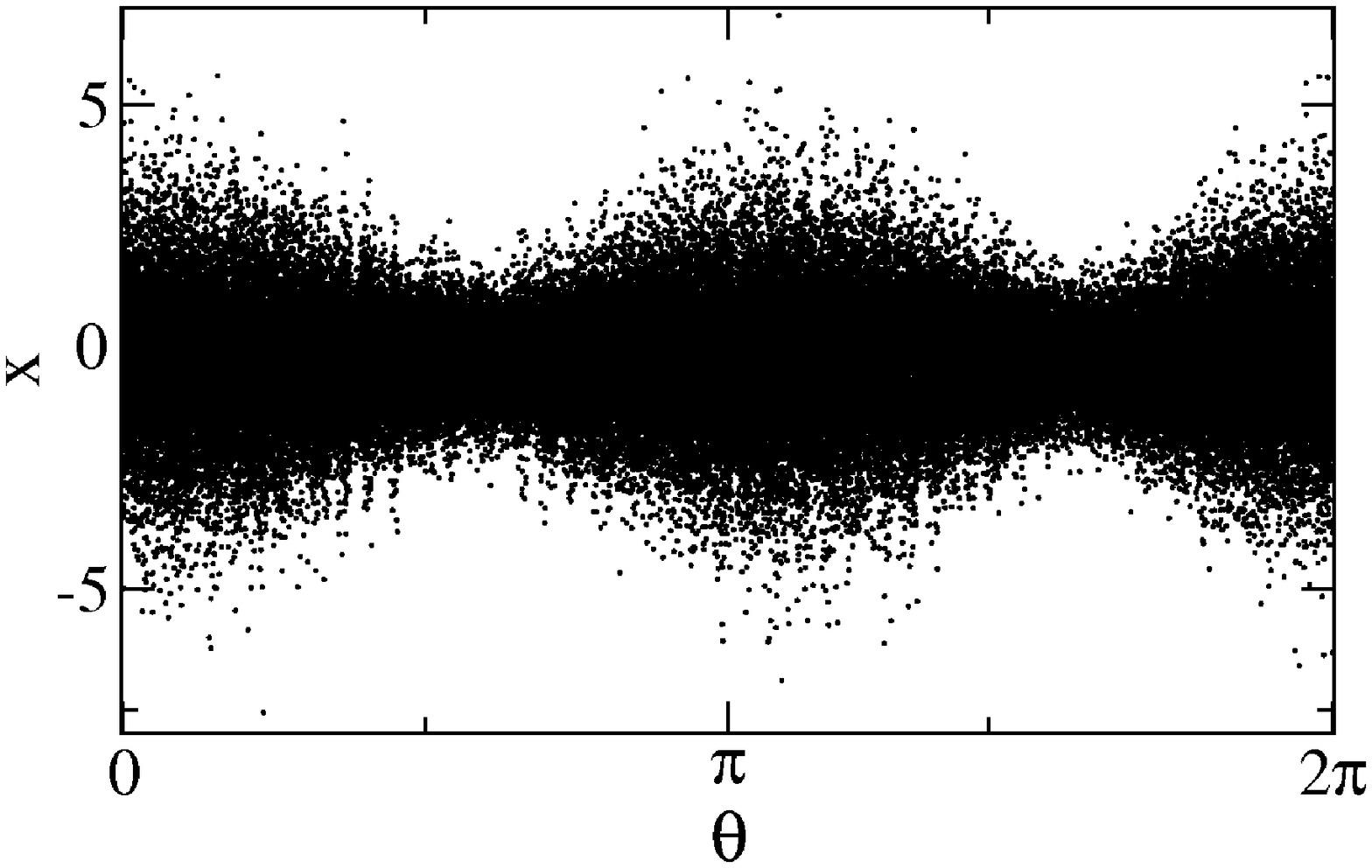}}
\caption{\label{fig:data} Raw homodyne data for the vacuum state 
(top) and for the OPO output close to threshold (bottom).}
\end{figure}\par
Data sampling is moved away from the laser carrier frequency by mixing
the homodyne current with a sinusoidal signal of frequency $ \Omega=3$
MHz. The resulting current is low--pass filtered with a bandwidth $B=1$
MHz, and eventually sampled by a digital acquisition PC based module
(Gage 14100) able to acquire up to 1M--points per run with 14 bits
resolution. For each state, 1~048~308 quadratures values were measured.
\par
Measured quadrature values normalized with respect to vacuum
fluctuations are shown in Fig.~\ref{fig:data}.  To facilitate our
analysis the data were grouped into $31$ phase bins prior to
reconstruction.  The purpose of the experiment was to perform Gaussian
tomography and compare the results with that of the standard homodyne
tomography \cite{objectivetomo} where no assumption about the
Gaussianity of the measured state were made. The comparison of the
reconstructed states may be seen as a test of Gaussian behavior.  Such
an approach is superior to a test based on evaluating certain quadrature
moments due to the fact that all the relevant moments are encoded in the
reconstructed state. 
\par
The Gaussian analysis starts with evaluating quantities 
$y_h=\sum_k x_{h,j}^2$ for each phase bin $\theta_h$, $h=1,31$.
These are used to construct operator $R$ 
appearing in  the extremal equation for covariance matrix Eq.~\eqref{ext-eqG}.
The following covariance matrices were found for measured vacuum 
(V) and OPO (O) data:
\begin{equation}\label{gaussrec}
G_\text{V}=\left( 
\begin{array}{cc}
0.50 & 0.00\\
0.00 & 0.50
\end{array}
\right),\,\,
G_\text{O}=\left( 
\begin{array}{cc}
2.38 & -0.53\\
-0.53 & 0.55
\end{array}
\right).
\end{equation}
Matrix $G_\text{O}$ indicates that the reconstructed Wigner function of
the OPO state is slightly rotated with respect to the $x$-$p$ axes,
which are defined by the the phase of the local oscillator.  The
squeezed nature of the input signal is revealed by diagonalizing matrix
$G_\text{O}$. With that variances of $0.40$ and $2.53$ are found for the
squeezed and anti-squeezed quadratures, respectively.  Notice that the
reconstructed state is not a minimum-uncertainty state, $\text{Det}\,
G_\text{O}=1.02 \gg 1/4$ and so the reconstructed covariance matrix is
located well \textit{inside} the set of physical covariance matrices,
$G_\text{O}>\Omega$.  To summarize the Gaussian analysis, a squeezed
Gaussian state is inferred from experimental data whose anti-squeezed
quadrature has been contaminated by some classical excess noise.
Now it is interesting to compare the result of Gaussian analysis with
that of the generic quantum tomography applied to the same data set. In
this case, no assumption is made about the Gaussian character of the
measured signal and the reconstruction is done on an $N$-dimensional
truncated Fock space using the maximum-likelihood technique
\cite{objectivetomo}.  Data have been grouped into $31$ phase
bins.  In addition, detections within each phase bin have been divided into
$31$ quadrature bins. The set of quadrature projectors corresponding to
the $31\times 31=961$ pairs of phase and quadrature values comprises a
POVM describing the tomography scheme.  Prior to reconstruction, the
dimension $N$ of the reconstruction space must be chosen. Notice that
this parameter has no analogy in the Gaussian analysis above. 
\begin{figure}[h!]
\centerline{\includegraphics[width=0.8\columnwidth]{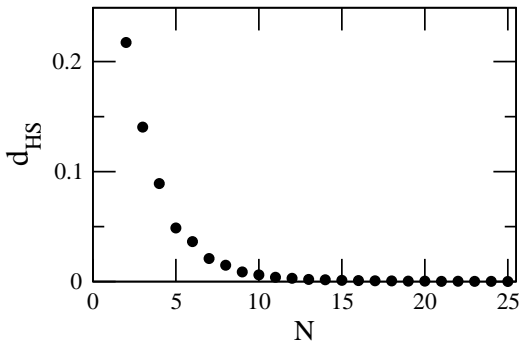}}
\caption{\label{fig:hs} 
Hilbert-Schmidt distances of state reconstructions 
done in $N$-dimensional Fock spaces with respect to the reference 
state reconstructed in dimension $N=30$.}
\end{figure}\par
\begin{figure*}
\centerline{\includegraphics[width=0.85\textwidth]{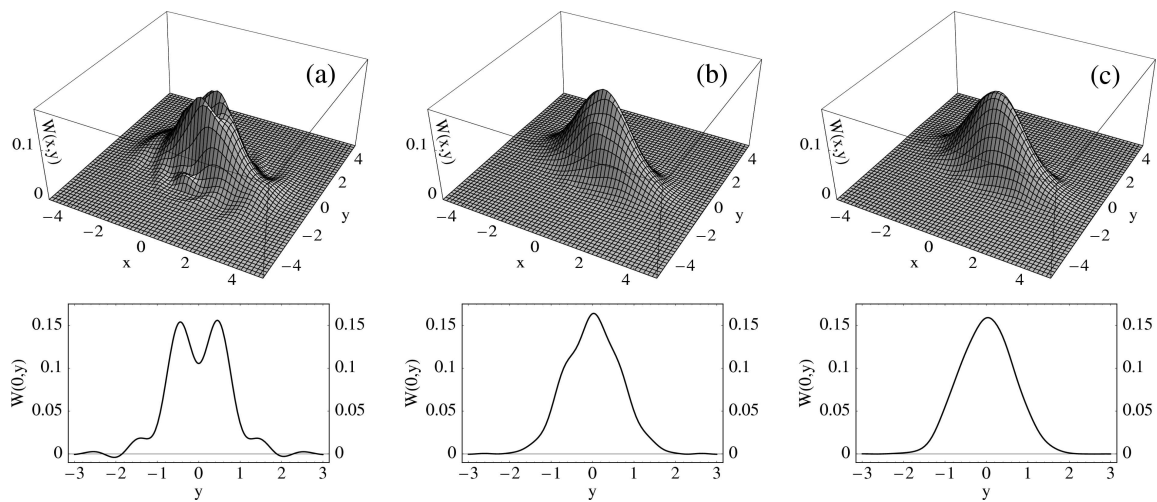}}
\caption{\label{fig:nongauss} Wigner functions
(upper row) and cuts along the $y$ axis (bottom row)
of the OPO state reconstructed in Fock spaces of dimensions
$N=8$ (a), $N=15$ (b), and $N=25$ (c) respectively.}
\end{figure*}

In order to see the effects of the truncation on the reconstructed state, 
the procedure has been repeated for different values of $N$.
The convergence of the generic reconstruction with growing
reconstruction space is shown in Fig.~\ref{fig:hs}, where the
Hilbert-Schmidt distance $d_\text{HS}=\text{Tr}\{(\rho_N-\rho_{30})^2\}$
between a reconstruction from a truncated space and a reference
reconstruction from a large space is plotted as a function of $N$. This
figure indicates that from $N=10$ on the reconstructed state does not
change significantly when the reconstruction space is enlarged.  This
conclusion, however, turns out to be incorrect when one is interested in
qualitative, rather than quantitative properties of the reconstructed
state.  For instance, the shape of the reconstructed Wigner function may
help to classify the signal as being Gaussian or non-Gaussian.
Similarly,  the presence of negative values of $W$ may be taken as a
sign of non-classical behavior.
In turn, making qualitative statements from quantitative results of
tomography may be delicate and challenging. This is illustrated in
Fig.~\ref{fig:nongauss}, where we report the Wigner function and the
corresponding y-cut for different values of the truncation dimension for
the reconstruction in the Fock space.  Notice that what may seem as a
sign of non-classical interference in Fig.~\ref{fig:nongauss}(a) is
actually an artifact due to small value of the reconstruction dimension.
Remarkably, this happens even though the inspection of the elements of
$\rho_{15}$ and $\rho_{25}$ in the Fock basis shows negligible
differences (see also Fig.~\ref{fig:hs}), since these tiny differences get
amplified when switching to Wigner representation. Also plots in
Fig.~\ref{fig:nongauss}(b) distinctly shows non-Gaussian shapes and the
artifacts are present until dimension of $N\approx 25$ is reached.  In
this region the generic reconstruction starts to be consistent with the
Gaussian results yielding very similar values of quadrature variances.
More than $600$ free parameters must be introduced in the standard
tomography to achieve the quality of the simple Gaussian fit of OPO
data!
\par
There remains a question whether the deviations from Gaussian character
that may be observed in the generic reconstruction even for large
dimensions $N$ are real features of the measured state or not.  This
possibility may be confirmed or discarded by analyzing the scores
(likelihoods) achieved by the Gaussian and generic reconstructions
respectively.  Here, a direct comparison is not possible due to
different natures of involved data. Instead, one may determine how much
the quality of the fit degrades by replacing vacuum data with the more
noisy OPO signal.  For Gaussian analysis we find that the log-likelihood
of the OPO reconstruction is smaller by a factor $3.66$ compared to the
vacuum reconstruction. Standard tomography yields a smaller factor of
$3.16$, which indicates that a better fit of the OPO signal generated
near the threshold is obtained by abandoning the Gaussian hypothesis.
\par
\begin{figure}
\centerline{
\includegraphics[width=0.85\columnwidth]{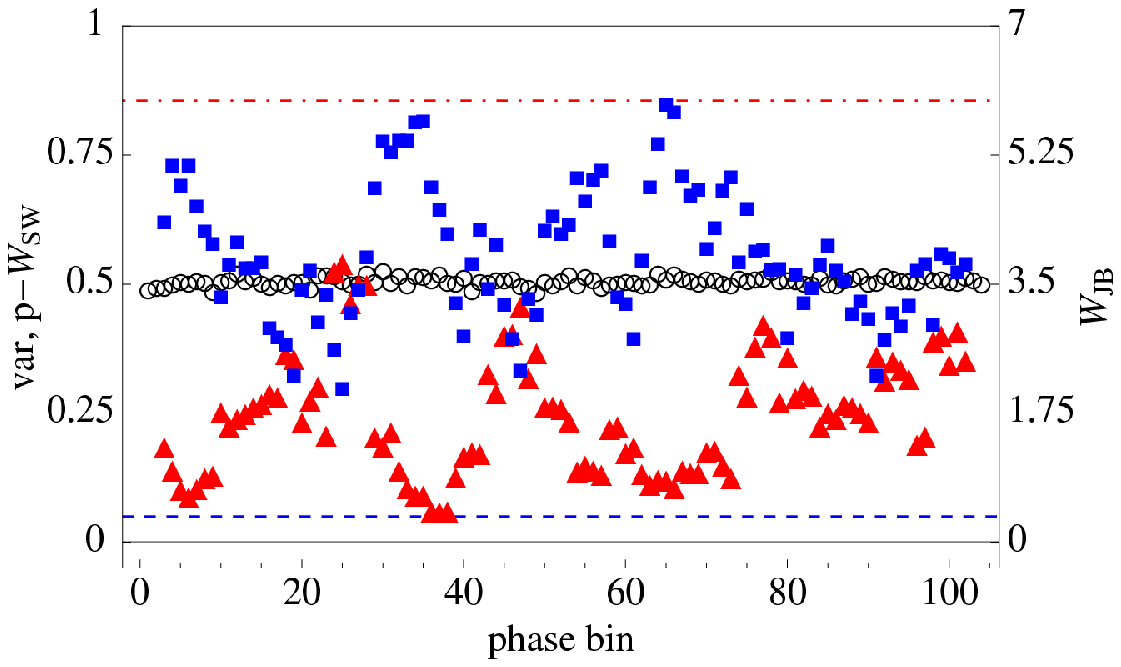}}
\vspace{0.3cm}
\centerline{\includegraphics[width=0.85\columnwidth]{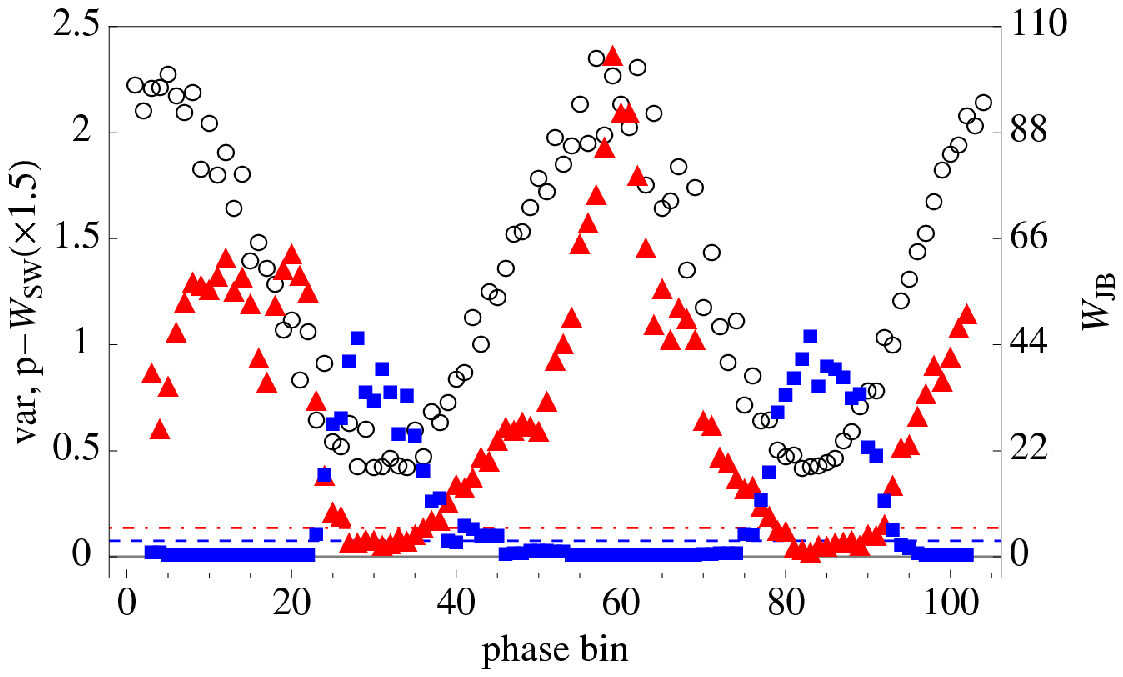}}
\caption{\label{f:NormTest} 
(Color online) Variance (circles), $W_{\rm JB}$ (triangles) and the $p$-value of
$W_{\rm SW}$ ($p$-$W_{\rm SW}$, squares) associated with the homodyne 
data of Figs.~\ref{fig:data}. Top: vacuum data. Bottom: OPO close to
threshold data (here $p$-$W_{\rm SW}$ has been magnified by a factor 
$\times 1.5$). 
The dashed line is the threshold value $0.05$ for $p$-$W_{\rm SW}$
($\times 1.5$ in the bottom plot); the dot-dashed line is the
threshold value $5.99$ for $W_{\rm JB}$. If $W_{\rm JB} \ge 5.99$ and/or
$p$-$W_{\rm SW} \le 0.05$, then the null hypothesis (Gaussian states) is 
rejected. See the text for details. 
Results indicate that vacuum data (top) are consistent with the
Gaussian hypothesis whereas for data coming from the OPO close to 
threshold the Gaussian hypothesis should be rejected.}
\end{figure}

In order to test  the null hypothesis for our samples 
(i.e., data normally distributed) we grouped homodyne data 
into phase bins made of  $10\,000$ outcomes each and applied 
``normality tests'' to both the samples of Figs.~\ref{fig:data}. 
In particular, we have used the Jarque-Bera (JB)
\cite{JB:80} and the Shapiro-Wilk (SW) normality tests \cite{SW:65}.
The JB test is based on both the skewness $S$ and 
the kurtosis (excess) $K$ of the sample, {\em i.e}
\begin{equation}
S=\frac{\mu_3}{(\mu_2)^{3/2}}\,,\quad
K=\frac{\mu_4}{(\mu_2)^{2}}-3\,,
\end{equation}
where $\mu_k \equiv \frac{1}{N}\sum_{h=1}^{N} (x_h-\overline{x})^k$ is the
$k$-th central moment, $N$ is the number of data and
$\overline{x}$ their mean. The JB statistics is then defined as:
\begin{equation}
W_{\rm JB} = \frac{N}{6}\left( S^2 + \frac14 K^2\right).
\end{equation}
Since $W_{\rm JB}$ has an asymptotic $\chi^2$ distribution with
two degrees of freedom \cite{JB:80}, one can use it to test
the null hypothesis of normality. In particular, by choosing a
significance level $0.05$, one rejects the null hypothesis if
$W_{\rm JB} > 5.99$. The test statistics $W_{\rm SW}$ of the SW test
reads as follows:
\begin{equation}
W_{\rm SW} =
\frac{\left(\sum_{h=1}^{N} a_h x_{(h)}\right)^2}
{\sum_{h=1}^{N}(x_h-\overline{x})^2},
\end{equation}
where $x_{(h)}$ are the ordered sample values ($x_{(h)}$ is the $h$-th
smallest value) and $a_h$ are constants generated from the means,
variances and covariances of the order statistics of a sample of
size $N$ from a normal distribution \cite{SW:65}. In this case one
rejects the null hypothesis of normality within a significance interval
of $0.05$, if $p$-$W_{\rm SW} \le 0.05$, where the  $p$-$W_{\rm SW}$ is
the $p$-value of $W_{\rm SW}$, {\em i.e.}
the probability of obtaining a result at least as extreme as the one
that was actually observed, given that the Gaussian hypothesis is true.
The two test statistics are shown in Fig.~\ref{f:NormTest},
where we plotted the variance, $W_{\rm JB}$ and $p$-$W_{\rm SW}$
associated with the data of Figs.~\ref{fig:data}. 
As one can see, the vacuum data are normally
distributed, whereas those coming from the OPO close to threshold show 
clear deviations from the Gaussian behavior (see also \cite{solimeno}), 
thus justifying the rejection of Gaussian hypothesis.
\section{Conclusions}
\label{outro}
In this paper, upon exploiting the formal analogies between the
description of Gaussian states and that of finite-dimensional quantum
states, we have proposed and demonstrated a simple and efficient
method for the reconstruction of Gaussian states.  The method
have been tested numerically and applied to the reconstruction of
the quantum state of the signal generated by an optical
parametric oscillator. As a matter of fact, under the Gaussian
hypothesis there is a drastic reduction in the number of parameters
needed to characterize a quantum state, and this leads to a
robust reconstruction technique. Indeed, our results indicate
that on Gaussian and near Gaussian states the proposed method
provide a more stable and reliable reconstructions than standard
tomography techniques. We have also shown that putting together
results coming from the Gaussian method with those coming from
standard tomography, we obtain a reliable tool to detect the
non-Gaussian character of optical signals.  We conclude that the
present method provides a reliable and robust tool for the
characterization of quantum states which may be of interest for a
rather broad community working in continuous-variable quantum
information processing and quantum state reconstruction.
\acknowledgments
The authors gratefully acknowledge the partial financial support
by projects of the Czech Grant Agency No. 202/06/307 and Czech
Ministry of Education ``Measurement and Information in Optics''
MSM 6198959213 and by the CNR-CNISM convention.
SO and MGAP thank Alberto Barchielli for useful suggestions
about test of Gaussianity.

\end{document}